\documentstyle[12pt]{article}
\begin{document}
\begin{center}

\title{Berry Phase in Generalized\\
            Chiral $QED_2$ \thanks{Extended version of a talk given at
XXI International Colloquium on Group-Theoretical Methods in
Physics, Goslar, Germany, 15-20 July, 1996}}

\author{{\bf Fuad M. Saradzhev} \thanks{Address till October 19, 1996:
ICTP, Trieste, Italy; e-mail: fuad@ictp.trieste.it}\\
{\it Institute of Physics, Academy of Sciences of 
Azerbaijan},\\
{\it Huseyn Javid pr. 33, 370143 Baku, AZERBAIJAN }
\thanks{Permanent address; e-mail: physic@lan.ab.az}}

\maketitle
\begin{flushleft}

{\bf ABSTRACT}

\rm
We consider the generalized chiral $QED_2$ on $S^1$ with a $U(1)$ gauge
field coupled with different charges to both chiral components of a
fermionic field. Using the adiabatic approximation we calculate the
Berry phase and the corresponding $U(1)$ connection
and curvature for the vacuum and many particle Fock states . We show that the 
nonvanishing vacuum Berry phase is 
associated with a projective representation of the local gauge symmetry group and
contributes to the effective action of the model.\\
\end{flushleft}
\end{center}

\newpage
\section{Introduction}
\label{sec: intro}

Gauge models with anomaly are interesting from different points of view.
First, there is a problem of consistent quantization for these models.
Due to anomaly some constraints change their nature after quantization:
instead of being  first-class constraints, they turn into second-class
ones. A consistent canonical   quantization scheme clearly should take
into account such a change \cite{jack85}-\cite{sarad91}.

Next is a problem of the relativistic invariance. It is known that in the
physical sector where the local gauge invariance holds the relativistic
invariance is broken for some anomalous models, namely the chiral
Schwinger model (CSM) and chiral $QCD_2$ \cite{niemi86}-\cite{sarad96}. 
For both models the Poincare
algebra commutation relations breaking term  can be constructed
explicitly \cite{sarad96}.

In the present paper we address ourselves to another aspect of anomalous
models: the Berry phase and its connection  to anomaly. A common topological
nature of the Berry phase, or more generally quantum holonomy, and gauge
anomalies was noted in \cite{alva85},\cite{niemi85}. The former was shown 
to be crucial in the hamiltonian interpretation of anomalies.

We consider a general version of the CSM with a ${\rm U}(1)$ gauge field
coupled with different charges to both chiral components of a fermionic
field. The non-anomalous Schwinger model (SM) where these charges are
equal is a special case of the generalized CSM. This will allow us
to see any distinction between the models with and without anomaly.

We suppose that space is a circle of length ${\rm L}$,
$-\frac{\rm L}{2} \leq  x < \frac{\rm L}{2}$, so space-time manifold
is a cylinder ${\rm S}^1 \otimes {\rm R}^1$. We work in the temporal
gauge $A_0=0$ and use the system of units where $c=1$. Only matter
fields are quantized, while $A_1$ is handled as a classical
background field. Our aim is to calculate the Berry phase and the
corresponding ${\rm U}(1)$ connection and curvature for the fermionic
Fock vacuum as well as for many particle states constructed over the
vacuum and to show explicitly a connection between the nonvanishing
vacuum Berry phase and anomaly.

Our paper is organized as follows. In  Sect.~\ref{sec: quant}, we
apply first and second quantization to the matter fields and obtain
the second quantized fermionic Hamiltonian. We define the Fock
vacuum and construct many particle Fock states over the vacuum. We
use a particle-hole interpretation for these states.

In Sect.~\ref{sec: berry} , we first derive a general formula for
the Berry phase and then calculate it for the vacuum  and many
particle states. We show that for all Fock states the Berry phase
vanishes in the case of models without anomaly. We discuss a connection
between the nonvanishing vacuum Berry phase, anomaly and effective
action of the model.

Our conclusions are in Sect.~\ref{sec: con}.

\newpage
\section{Quantization of matter fields}
\label{sec: quant}

The Lagrangian density of the generalized CSM  is
\begin{equation}
{\cal L} = - {\frac{1}{4}} {\rm F}_{\mu \nu} {\rm F}^{\mu \nu} +
\bar{\psi} i {\hbar} {\gamma}^{\mu} {\partial}_{\mu} \psi +
e_{+} {\hbar} \bar{\psi}_{+} {\gamma}^{\mu} {\psi_{+}} A_{\mu} +
e_{-} {\hbar} \bar{\psi}_{-} {\gamma}^{\mu} {\psi_{-}} A_{\mu} ,
\label{eq: odin}
\end{equation}
where ${\rm F}_{\mu \nu}= \partial_{\mu} A_{\nu} - \partial_{\nu} A_{\mu}$ ,
$(\mu, \nu) = \overline{0,1}$ , $\gamma^{0}={\sigma}_1$,
${\gamma}^{1}=-i{\sigma}_2$, ${\gamma}^0 {\gamma}^1={\gamma}^5=
{\sigma}_3$, ${\sigma}_i (i=\overline{1,3})$ are Pauli matrices.
The field $\psi$ is $2$--component Dirac spinor, $\bar{\psi} =
\psi^{\dagger} \gamma^0$ and $\psi_{\pm}=\frac{1}{2} (1 \pm \gamma^5)
\psi$.

In the temporal gauge $A_0=0$, the Hamiltonian density is
\begin{equation}
{\cal H}  =  \frac{1}{2}{\rm E}^2 + {\cal H}_{+} + {\cal H}_{-},
\label{eq: dva}
\end{equation}
with ${\rm E}$ momentum canonically conjugate to $A_1$, and
\[
{\cal H}_{\pm}  \equiv  \hbar \psi_{\pm}^{\dagger} d_{\pm} \psi_{\pm} =
\mp \hbar \psi_{\pm}^{\dagger}(i{\partial}_{1}+e_{\pm}A_1)\psi_{\pm}.
\]

On the circle boundary conditions for the fields must be specified.
We impose the periodic ones
\begin{eqnarray}
{A_1} (- \frac{\rm L}{2}) & = & {A_1} (\frac{\rm L}{2}) \nonumber \\
{\psi_{\pm}} (- \frac{\rm L}{2}) & = & {\psi_{\pm}} (\frac{\rm L}{2}).
\label{eq: tri}
\end{eqnarray}
The Lagrangian and Hamiltonian densities
are invariant under local time-independent gauge 
transformations
\begin{eqnarray*}
A_1 & \rightarrow & A_1 + {\partial}_{1} \lambda,\\
\psi_{\pm} & \rightarrow & \exp\{ie_{\pm} \lambda\} \psi_{\pm},
\end{eqnarray*}
$\lambda$ being a gauge function.

For arbitrary $e_{+},e_{-}$, the gauge transformations do not respect
the boundary conditions ~\ref{eq: tri}.
The gauge transformations compatible with the boundary conditions
must be either of the form
\[
\lambda (\frac{\rm L}{2})=\lambda (- \frac{\rm L}{2}) +
{\frac{2\pi}{e_{+}}}n,
\hspace{5 mm}
{\rm n} \in \cal Z.
\]
with $e_{+} \neq 0$ and
\begin{equation}
\frac{e_{-}}{e_{+}} = {\rm N},
\hspace{5 mm}
{\rm N} \in \cal Z,
\label{eq: cet}
\end{equation}
or of the form
\[
\lambda(\frac{\rm L}{2}) = \lambda(-\frac{\rm L}{2}) +
\frac{2\pi}{e_{-}} n ,
\hspace{5 mm}
{\rm n} \in \cal Z,
\]
with $e_{-} \neq 0$ and
\begin{equation}
\frac{e_{+}}{e_{-}} = \bar{\rm N},
\hspace{5 mm}
\bar{\rm N} \in \cal Z.
\label{eq: pet}
\end{equation}
Eqs. ~\ref{eq: cet} and ~\ref{eq: pet} imply a quantization condition
for the charges. Without loss of generality, we choose ~\ref{eq: cet}.
For ${\rm N}=1$, $e_{-}=e_{+}$ and we have the standard Schwinger model.
For ${\rm N}=0$, we get the model in which only the positive chirality
component of the Dirac field is coupled to the gauge field.

We see that the gauge transformations under consideration are divided
into topological classes characterized by the integer $n$. If
$\lambda(\frac{\rm L}{2}) = \lambda(-\frac{\rm L}{2})$, then the
gauge transformation is topologically trivial and belongs to the
$n=0$ class. If $n \neq 0$ it is nontrivial and has winding number $n$.

The eigenfunctions and the eigenvalues of the first quantized
fermionic Hamiltonians are
\[
d_{\pm} \langle x|n;{\pm} \rangle = \pm \varepsilon_{n,{\pm }}
\langle x|n;{\pm } \rangle ,
\]
where
\[
\langle x|n;{\pm } \rangle = \frac{1}{\sqrt {\rm L}}
\exp\{ie_{\pm} \int_{-{\rm L}/2}^{x} dz{A_1}(z) +
i\varepsilon_{n,{\pm}} \cdot x\},
\]
\[
\varepsilon_{n,{\pm }} = \frac{2\pi}{\rm L}
(n - \frac{e_{\pm}b{\rm L}}{2\pi}).
\]
We see that the spectrum of the eigenvalues depends on the zero
mode of the gauge field:
\[
b \equiv \frac{1}{\rm L} \int_{-{\rm L}/2}^{{\rm L}/2} dx
A_1(x,t).
\]
For $\frac{e_{+}b{\rm L}}{2\pi}={\rm integer}$, the spectrum contains
the zero energy level. As $b$ increases from $0$ to
$\frac{2\pi}{e_{+}{\rm L}}$, the energies of
$\varepsilon_{n,+}$ decrease by $\frac{2\pi}{\rm L}$, while the energies
of $(-\varepsilon_{n,-})$ increase by $\frac{2\pi}{\rm L} {\rm N}$.
Some of energy levels change sign. However, the spectrum at the
configurations $b=0$ and $b=\frac{2\pi}{e_{+}{\rm L}}$
is the same, namely, the integers, as it must be since these gauge-field
configurations are gauge-equivalent. In what follows, we
will use separately the integer and fractional parts of
$\frac{e_{\pm}b{\rm L}}{2\pi}$, denoting them as
$[\frac{e_{\pm}b{\rm L}}{2\pi}]$ as $\{\frac{e_{\pm}b{\rm L}}{2\pi}\}$ 
correspondingly.

Now we introduce the second quantized right-handed and
left-handed Dirac fields. For the moment, we will assume that $d_{\pm}$
do not have zero eigenvalues. At time $t=0$, in terms of the
eigenfunctions of the first quantized fermionic Hamiltonians the second
quantized ($\zeta$--function regulated) fields have the expansion
\cite{niese86} :
\[
\psi_{+}^s (x) = \sum_{n \in \cal Z} a_n \langle x|n;{+} \rangle
|\lambda \varepsilon_{n,+}|^{-s/2},
\]
\begin{equation}
\psi_{-}^s (x) = \sum_{n \in \cal Z} b_n \langle x|n;{-} \rangle
|\lambda \varepsilon_{n,-}|^{-s/2}.
\label{eq: vosem}
\end{equation}
Here $\lambda$ is an arbitrary constant with dimension of length
which is necessary to make $\lambda \varepsilon_{n,\pm}$ dimensionless,
while $a_n, a_n^{\dagger}$ and $b_n, b_n^{\dagger}$ are correspondingly
right-handed and left-handed fermionic creation and annihilation
operators which fulfil the commutation relations
\[
[a_n , a_m^{\dagger}]_{+} = [b_n , b_n^{\dagger}]_{+} =\delta_{m,n} .
\]
For $\psi_{\pm }^{s} (x)$, the equal time anticommutators are
\begin{equation}
[\psi_{\pm}^{s}(x) , \psi_{\pm}^{\dagger s}(y)]_{+}=\zeta_{\pm} (s,x,y),
\label{eq: devet}
\end{equation}
with all other anticommutators vanishing, where
\[
\zeta_{\pm} (s,x,y) \equiv \sum_{n \in \cal Z} \langle x|n;{\pm} \rangle
\langle n;{\pm}|y \rangle |\lambda \varepsilon_{n,\pm}|^{-s},
\]
$s$ being large and positive. In the limit, when the regulator
is removed, i.e. $s=0$, $\zeta_{\pm}(s=0,x,y) = \delta(x-y)$ and
Eq.~\ref{eq: devet} takes the standard form.

The vacuum state of the second quantized fermionic Hamiltonian
\[
|{\rm vac};A \rangle = |{\rm vac};A;+ \rangle \otimes
|{\rm vac};A;- \rangle
\]
is defined such that all negative energy
levels are filled and the others are empty:
\begin{eqnarray}
a_n|{\rm vac};A;+\rangle =0 & {\rm for} & n>[\frac{e_{+}b{\rm L}}{2\pi}],
\nonumber \\
a_n^{\dagger} |{\rm vac};A;+ \rangle =0 & {\rm for} & n \leq
[\frac{e_{+}b{\rm L}}{2\pi}],
\label{eq: deset}
\end{eqnarray}
and
\begin{eqnarray}
b_n|{\rm vac};A;-\rangle =0 & {\rm for} & n \leq
[\frac{e_{-}b{\rm L}}{2\pi}], \nonumber \\
b_n^{\dagger} |{\rm vac};A;- \rangle =0 & {\rm for} & n >
[\frac{e_{-}b{\rm L}}{2\pi}].
\label{eq: odinodin}
\end{eqnarray}
In other words, in the positive chirality vacuum all the levels
with energy lower than ${\varepsilon}_{[\frac{e_{+}b{\rm L}}
{2\pi}]+1,+}$ and in the negative chirality one all the levels
with energy lower than $(-{\varepsilon}_{[\frac{e_{-}b{\rm L}}
{2\pi}],-})$ are filled:
\begin{eqnarray*}
|{\rm vac}; A;+ \rangle & = & \prod_{n=\infty}^{[\frac{e_{+}b
{\rm L}}{2\pi}]} a_m^{\dagger} |0;+ \rangle, \\
|{\rm vac}; A;- \rangle & = & \prod_{n=[\frac{e_{-}b{\rm L}}
{2\pi}]+1}^{+\infty} b_n^{\dagger} |0;- \rangle, \\
\end{eqnarray*}
where $|0 \rangle = |0,+ \rangle \otimes |0,- \rangle$ is the state
of "nothing" with all the energy levels empty.

The Fermi surfaces which are defined to lie halfway between the highest
filled and lowest empty levels are
\[
{\varepsilon}_{\pm}^{\rm F} = \pm \frac{2\pi}{\rm L}
(\frac{1}{2} - \{\frac{e_{\pm}b{\rm L}}{2\pi}\}).
\]
For $e_{+}=e_{-}$, ${\varepsilon}_{+}^{\rm F}=-{\varepsilon}_{-}^{\rm F}$.

Next we define the fermionic parts of the second-quantized Hamiltonian as
\[
\hat{\rm H}_{\pm}^s = \int_{-{\rm L}/2}^{{\rm L}/2} dx
\hat{\cal H}_{\pm}^s(x)= \frac{1}{2} \hbar \int_{-{\rm L}/2}^{{\rm L}/2}dx
 (\psi_{\pm}^{\dagger s} d_{\pm} \psi_{\pm}^s
- \psi_{\pm}^s d_{\pm}^{\star} \psi_{\pm}^{\dagger s}).
\]
Substituting ~\ref{eq: vosem} into this expression, we get
\begin{equation}
\hat{\rm H}_{\pm} =  \hat{\rm H}_{0,\pm}  \mp
e_{\pm} b \hbar :\rho_{\pm}(0): + {\hbar} \frac{\rm L}{4\pi}  
({\varepsilon}_{\pm}^{\rm F})^2, 
\label{eq: hamil}
\end{equation}
where double dots indicate normal ordering with respect to
$|{\rm vac},A \rangle$ ,
\begin{eqnarray*}
\hat{\rm H}_{0,+} & = & \hbar \frac{2 \pi}{\rm L} \lim_{s \to 0}
\{ \sum_{k >[\frac{e_{+}b{\rm L}}{2 \pi}]} k a_k^{\dagger} a_k
|\lambda \varepsilon_{k,+}|^{-s}  -  \sum_{k \leq [\frac{e_{+}b{\rm L}}
{2 \pi}]} k a_k a_k^{\dagger} |\lambda \varepsilon_{k,+}|^{-s} \},\\
\hat{\rm H}_{0,-} & = & \hbar \frac{2 \pi}{\rm L} \lim_{s \to 0}
\{ \sum_{k>[\frac{e_{-}b{\rm L}}{2 \pi}]} k b_{k} b_{k}^{\dagger}
|\lambda \varepsilon_{k,-}|^{-s} - \sum_{k \leq [\frac{e_{-}b{\rm L}}
{2 \pi}]} k b_{k}^{\dagger} b_{k} |\lambda \varepsilon_{k,-}|^{-s} \}
\end{eqnarray*}
are free fermionic Hamiltonians, and
\begin{eqnarray*}
:\rho_{+} (0): & = & \lim_{s \to 0} \{ \sum_{k >[\frac{e_{+}b{\rm L}}
{2 \pi}]} a_k^{\dagger} a_k |\lambda \varepsilon_{k,+}|^{-s}  -
\sum_{k \leq [\frac{e_{+}b{\rm L}}{2 \pi}]} a_k a_k^{\dagger}
|\lambda \varepsilon_{k,+}|^{-s} \}, \\
:\rho_{-} (0): & = & \lim_{s \to 0} \{ \sum_{k \leq [\frac{e_{-}b{\rm L}}
{2 \pi}]} b_{k}^{\dagger}  b_{k} |\lambda \varepsilon_{k,-}|^{-s} -
\sum_{k>[\frac{e_{-}b{\rm L}}{2 \pi}]} b_{k} b_{k}^{\dagger}
|\lambda \varepsilon_{k,-}|^{-s} \} 
\end{eqnarray*}
are charge operators for the positive and negative chirality fermion
fields respectively. The fermion momentum operators constructed
analogously are
\[
\hat{\rm P}_{\pm} = \hat{\rm H}_{0,\pm}.
\]
The operators $:\hat{\rm H}_{\pm}:$, $:\rho_{\pm}(0):$ 
and $\hat{\rm P}_{\pm}$ are
well defined when acting on finitely excited states which have only a
finite number of excitations relative to the Fock vacuum.

For the vacuum state,
\[
:\hat{\rm H}_{\pm}:|{\rm vac}; A;\pm \rangle =
:{\rho}_{\pm}(0):|{\rm vac}; A;\pm \rangle =0.
\]
Due to the normal ordering, the energy of the vacuum which is at the
same time the ground state of the fermionic Hamiltonians turns out
to be equal to zero ( we neglect an infinite energy of the filled
levels below the Fermi surfaces ${\varepsilon}_{\pm}^{\rm F}$).
The vacuum state can be considered also as a state of the zero charge.

Any other state of the same charge will have some of the levels above
${\varepsilon}_{+}^{\rm F}$ (${\varepsilon}_{-}^{\rm F}$) occupied
and some levels below ${\varepsilon}_{+}^{\rm F}$ (${\varepsilon}_{-}
^{\rm F}$) unoccupied. It is convenient to use the vacuum state
$|{\rm vac}; A \rangle$ as a reference, describing the removal
of a particle of positive (negative) chirality from one of the levels
below ${\varepsilon}_{+}^{\rm F}$ (${\varepsilon}_{-}^{\rm F}$) as
the creation of a "hole" \cite{dirac64},\cite{feyn72}. 
Particles in the levels above
${\varepsilon}_{+}^{\rm F}$ (${\varepsilon}_{-}^{\rm F}$) are still
called particles. If a particle of positive (negative) chirality
is excited from the level $m$ below the Fermi surface to the level
$n$ above the Fermi surface, then we say that a hole of positive
chirality with energy $(-{\hbar}{\varepsilon}_{m,+})$ and
momentum $(-{\hbar}\frac{2\pi}{\rm L} m)$ ( or of negative chirality
with energy ${\hbar}{\varepsilon}_{m,-}$ and momentum
${\hbar}\frac{2\pi}{\rm L} m$) has been created as well as the
positive chirality particle with energy ${\hbar}{\varepsilon}_{n,+}$
and momentum ${\hbar}\frac{2\pi}{\rm L}n$ ( or the negative chirality
one with energy $(-{\hbar}{\varepsilon}_{n,-})$ and momentum
$(-{\hbar}\frac{2\pi}{\rm L}n)$ ). The operators $a_k (k \leq
[\frac{e_{+}b{\rm L}}{2\pi}])$ and $b_k (k>[\frac{e_{-}b{\rm L}}{2\pi}])$
behave like creation operators for the positive and negative chirality
holes correspondingly.

In the charge operator a hole counts as $-1$, so that, for example,
any state with one particle and one hole as well as the vacuum state
has vanishing charge.

The number of particles and holes of positive and negative chirality
outside the vacuum state is given by the operators
\begin{eqnarray*}
{\rm N}_{+} & = & \lim_{s \to 0} \{ \sum_{k>[\frac{e_{+}b{\rm L}}
{2\pi}]} a_k^{\dagger} a_k + \sum_{k \leq [\frac{e_{+}b{\rm L}}
{2\pi}]} a_k a_k^{\dagger} \} |{\lambda}{\varepsilon}_{k,+}|^{-s}, \\
{\rm N}_{-} & = & \lim_{s \to 0} \{ \sum_{k \leq [\frac{e_{-}b{\rm L}}
{2\pi}]} b_k^{\dagger} b_k + \sum_{k>[\frac{e_{-}b{\rm L}}{2\pi}]}
b_k b_k^{\dagger} \} |{\lambda}{\varepsilon}_{k,-}|^{-s},\\
\end{eqnarray*}
which count both particle and hole as $+1$.

Excited states are constructed by operating creation operators
on the vacuum. We start with $1$-particle states. Let us define the 
states $|m; A;\pm \rangle$ as follows
$$
|m; A;+ \rangle \equiv \left\{
\begin{array}{cc} 
a_m^{\dagger}|{\rm vac}; A;+ 
\rangle & {\rm for} \hspace{5 mm} m>[\frac{e_{+}b{\rm L}}{2\pi}], \\ 
a_m |{\rm vac}; A;+ 
\rangle & {\rm for} \hspace{5 mm} m \leq [\frac{e_{+}b{\rm L}}{2\pi}] 
\end{array}
\right.
$$
and
$$
|m; A;- \rangle \equiv 
\left\{ \begin{array}{cc} 
b_m^{\dagger} |{\rm vac}; A;- 
\rangle & {\rm for} \hspace{5 mm}  m \leq [\frac{e_{-}b{\rm L}}{2\pi}],\\ 
b_m |{\rm vac}; A;- \rangle & {\rm for} \hspace{5 mm} m>[\frac{e_{-}b{\rm 
L}}{2\pi}]. \end{array}
\right .
$$
The states $|m; A;\pm \rangle$ are orthonormalized,
\[
\langle m; A;\pm |n,  A; \pm \rangle = \delta_{mn},
\]
and fulfil the completeness relation
\[
\sum_{m \in \cal Z} |m; A;\pm \rangle \cdot
\langle m; A;\pm| =1.
\]
It is easily checked that
\begin{eqnarray*}
:\hat{\rm H}_{\pm}: |m; A;\pm \rangle & = & {\hbar}{\varepsilon}
_{m,\pm} |m; A;\pm \rangle, \\
\hat{\rm P}_{\pm} |m; A;\pm \rangle & = & {\hbar}\frac{2\pi}{\rm L}
m |m; A;\pm \rangle, \\
:{\rho}_{\pm}(0): |m; A;\pm \rangle & = & \pm |m; A;\pm \rangle
\hspace{5 mm}
{\rm for}
\hspace{5 mm}
m > [\frac{e_{\pm}b{\rm L}}{2\pi}]
\end{eqnarray*}
and
\begin{eqnarray*}
:\hat{\rm H}_{\pm}: |m; A;\pm \rangle & = & - {\hbar}{\varepsilon}
_{m,\pm} |m; A;\pm \rangle, \\
\hat{\rm P}_{\pm} |m; A;\pm \rangle & = & -{\hbar} \frac{2\pi}{\rm L}
m |m; A;\pm \rangle, \\
:{\rho}_{\pm}(0): |m; A;\pm \rangle & = & \mp |m; A;\pm \rangle
\hspace{5 mm}
{\rm for}
\hspace{5 mm}
m \leq [\frac{e_{\pm}b{\rm L}}{2\pi}].
\end{eqnarray*}
We see that $|m; A;+ \rangle$ is a state with one particle of
positive chirality with energy ${\hbar}{\varepsilon}_{m,+}$ and
momentum ${\hbar}\frac{2\pi}{\rm L} m$ for $m>[\frac{e_{+}b{\rm L}}
{2\pi}]$ or a state with one hole of the same chirality with energy
$(-{\hbar}{\varepsilon}_{m,+})$ and momentum $(-\hbar \frac{2\pi}{\rm L}
m)$ for $m \leq [\frac{e_{+}b{\rm L}}{2\pi}]$. The negative chirality
state $|m; A;- \rangle$ is a state with one particle with energy
$(-\hbar {\varepsilon}_{m,-})$ and momentum $(-\hbar \frac{2\pi}{\rm L}
m)$ for $m \leq [\frac{e_{-}b{\rm L}}{2\pi}]$ or a state with one hole
with energy $\hbar {\varepsilon}_{m,-}$ and momentum $\hbar 
\frac{2\pi}{\rm L} m$ for $m >[\frac{e_{-}b{\rm L}}{2\pi}]$. In any
case,
\[
{\rm N}_{\pm}|m; A;\pm \rangle = |m; A;\pm \rangle,
\]
that is why $|m; A;\pm \rangle$ are called $1$-particle states.

By applying $n$ creation operators to the vacuum states $|{\rm vac};
A;\pm \rangle$ we can get also $n$-particle states
\[
|m_1;m_2;...;m_n; A;\pm \rangle
\hspace{5 mm}
(m_1 < m_2 < ... <m_n),
\]
which are orthonormalized:
\[
\langle m_1;m_2;...;m_n; A;\pm |\overline{m}_1;\overline{m}_2;
...;\overline{m}_n; A;\pm \rangle =
{\delta}_{m_1 \overline{m}_1} {\delta}_{m_2 \overline{m}_2} ...
{\delta}_{m_n \overline{m}_n}.
\]
The completeness relation is written in the following form
\begin{equation}
\frac{1}{n!} \sum_{m_1 \in \cal Z} ... \sum_{m_n \in \cal Z}
|m_1;m_2;...;m_n; A;\pm \rangle \cdot
\langle m_1;m_2;...;m_n; A;\pm| =1.
\label{eq: polnota}
\end{equation}
Here the range of $m_i$ ($i=\overline{1,n}$) is not restricted by
the condition $(m_1<m_2<...<m_n)$, duplication of states being taken care
of by the $1/n!$ and the normalization. The $1$ on the right-hand side
of Eq.~\ref{eq: polnota} means the unit operator on the space of
$n$-particle states.

The case $n=0$ corresponds to the zero-particle states. They form a
one-dimensional space, all of whose elements are proportional to
the vacuum state.

The multiparticle Hilbert space is a direct sum of an infinite
sequence of the $n$-particle Hilbert spaces. The states of different
numbers of particles are defined to be orthogonal to each other.

The completeness relation in the multiparticle Hilbert space has the
form
\begin{equation}
\sum_{n=0}^{\infty} \frac{1}{n!} \sum_{m_1,m_2,...m_n \in \cal Z}
|m_1;m_2;...;m_n; A;\pm \rangle \cdot
\langle m_1;m_2;...;m_n; A;\pm| = 1,
\label{eq: plete}
\end{equation}
where "1" on the right-hand side means the unit operator on the
whole multiparticle space.

For $n$-particle states,
\[
:\hat{\rm H}_{\pm}: |m_1;m_2;...;m_n; A;\pm \rangle =
\hbar \sum_{k=1}^{n} {\varepsilon}_{m_k,\pm} \cdot {\rm sign}
({\varepsilon}_{m_k,\pm}) |m_1;m_2;...;m_n; A;\pm \rangle
\]
and
\[
:{\rho}_{\pm}(0): |m_1;m_2;...;m_n; A;\pm \rangle =
\pm \sum_{k=1}^{n} {\rm sign}({\varepsilon}_{m_k,\pm})
|m_1;m_2;...;m_n; A;\pm \rangle.
\]

\newpage
\section{Calculation of Berry phases}
\label{sec: berry}

In the adiabatic approach \cite{schiff68}-\cite{zwanz}, the 
dynamical
variables are divided into two sets, one which we call fast variables
and the other which we call slow variables. In our case, we treat the
fermions as fast variables and the gauge fields as slow variables.

Let ${\cal A}^1$ be a manifold of all static gauge field
configurations ${A_1}(x)$. On ${\cal A}^1$  a time-dependent
gauge field ${A_1}(x,t)$ corresponds to a path and a periodic gauge
field to a closed loop.

We consider the fermionic part of the second-quantized Hamiltonian
$:\hat{\rm H}_{\rm F}:=:\hat{\rm H}_{+}: + :\hat{\rm H}_{-}:$ 
which depends on $t$  through the background
gauge field $A_1$ and so changes very slowly with time. We consider
next the periodic gauge field ${A_1}(x,t) (0 \leq t <T)$ . After a
time $T$ the periodic field ${A_1}(x,t)$ returns to its original
value: ${A_1}(x,0) = {A_1}(x,T)$, so that $:\hat{\rm H}_{\pm}:(0)=
:\hat{\rm H}_{\pm}:(T)$ .

At each instant $t$ we define eigenstates for $:\hat{\rm H}_{\pm}:
(t)$ by
\[
:\hat{\rm H}_{\pm}:(t) |{\rm F}, A(t);\pm \rangle =
{\varepsilon}_{{\rm F},\pm}(t) |{\rm F}, A(t);\pm \rangle.
\]
The state $|{\rm F}=0, A(t);\pm \rangle \equiv |{\rm vac}; A(t);\pm \rangle$
is a ground state of $:\hat{\rm H}_{\pm}:(t)$ ,
\[
:\hat{\rm H}_{\pm}:(t) |{\rm vac}; A(t);\pm \rangle =0.
\]
The Fock states $|{\rm F}, A(t) \rangle = |{\rm F},A(t);+ \rangle
\otimes |{\rm F},A(t);- \rangle $ 
depend on $t$ only through
their implicit dependence on $A_1$. They are assumed to be
orthonormalized,
\[
\langle {\rm F^{\prime}}, A(t)|{\rm F}, A(t) \rangle =
\delta_{{\rm F},{\rm F^{\prime}}},
\]
and nondegenerate.

The time evolution of the wave function  of our system (fermions
in a background gauge field) is clearly governed by the Schrodinger
equation:
\[
i \hbar \frac{\partial \psi(t)}{\partial t} =
:\hat{\rm H}_{\rm F}:(t) \psi(t) .
\]
For each $t$, this wave function can be expanded in terms of the
"instantaneous" eigenstates $|{\rm F}, A(t) \rangle$ .

Let us choose ${\psi}_{\rm F}(0)=|{\rm F}, A(0) \rangle$, i.e.
the system is initially described by the eigenstate
$|{\rm F},A(0) \rangle$ . According to the adiabatic approximation,
if at $t=0$ our system starts in an stationary state $|{\rm F},A(0)
\rangle $ of $:\hat{\rm H}_{\rm F}:(0)$, then it will remain,
at any other instant of time $t$, in the corresponding eigenstate
$|{\rm F}, A(t) \rangle$ of the instantaneous Hamiltonian
$:\hat{\rm H}_{\rm F}:(t)$. In other words, in the adiabatic
approximation transitions to other eigenstates are neglected.

Thus, at some time $t$ later our system will be described up to
a phase by the same Fock state $|{\rm F}, A(t) \rangle $:
\[
\psi_{\rm F}(t) = {\rm C}_{\rm F}(t) \cdot |{\rm F},A(t) \rangle,
\]
where ${\rm C}_{\rm F}(t)$ is yet undetermined phase.

To find the phase, we insert $\psi_{\rm F}(t)$ into the
Schrodinger equation :
\[
\hbar \dot{\rm C}_{\rm F}(t) = -i {\rm C}_{\rm F}(t)
(\varepsilon_{{\rm F},+}(t) + \varepsilon_{{\rm F},-}(t)) 
- \hbar {\rm C}_{\rm F}(t)
\langle {\rm F},A(t)|\frac{\partial}{\partial t}|{\rm F},A(t) \rangle.
\]
Solving this equation, we get
\[
{\rm C}_{\rm F}(t) = \exp\{- \frac{i}{\hbar} \int_{0}^{t} d{t^{\prime}}
({\varepsilon}_{{\rm F},+}({t^{\prime}}) +
{\varepsilon}_{{\rm F},-}({t^{\prime}}) ) - \int_{0}^{t} d{t^{\prime}}
\langle {\rm F},A({t^{\prime}})|\frac{\partial}{\partial{t^{\prime}}}|
{\rm F},A({t^{\prime}}) \rangle \}.
\]
For $t=T$, $|{\rm F},A(T) \rangle =|{\rm F},A(0) \rangle$ ( the
instantaneous eigenfunctions are chosen to be periodic in time)
and
\[
{\psi}_{\rm F}(T) = \exp\{i {\gamma}_{\rm F}^{\rm dyn} +
i {\gamma}_{\rm F}^{\rm Berry} \}\cdot {\psi}_{\rm F}(0),
\]
where
\[ {\gamma}_{\rm F}^{\rm dyn} \equiv - \frac{1}{\hbar}
\int_{0}^{T} dt \cdot ({\varepsilon}_{{\rm F},+}(t) 
+ {\varepsilon}_{{\rm F},-}(t)),    
\]
while
\begin{equation}
{\gamma}_{\rm F}^{\rm Berry} = {\gamma}_{\rm F,+}^{\rm Berry} +
{\gamma}_{\rm F,-}^{\rm Berry},
\label{eq: summa}
\end{equation}
\[
{\gamma}_{{\rm F},\pm}^{\rm Berry} \equiv \int_{0}^{T} dt \int_{-{\rm L}/2}^
{{\rm L}/2} dx \dot{A_1}(x,t) \langle {\rm F},A(t);\pm|i \frac{\delta}
{\delta A_1(x,t)}|{\rm F},A(t);\pm \rangle
\]
is Berry's phase  \cite{berry84}.

If we define the $U(1)$ connections
\begin{equation}
{\cal A}_{{\rm F},\pm}(x,t) \equiv \langle {\rm F},A(t);\pm|i \frac{\delta}
{\delta A_1(x,t)}|{\rm F},A(t);\pm \rangle,
\label{eq: dvatri}
\end{equation}
then
\[
{\gamma}_{{\rm F},\pm}^{\rm Berry} = \int_{0}^{T} dt \int_{-{\rm L}/2}^
{{\rm L}/2} dx \dot{A}_1(x,t) {\cal A}_{{\rm F},\pm}(x,t).
\]
We see that upon parallel transport around a closed loop on
${\cal A}^1$ the Fock states $|{\rm F},A(t);\pm \rangle$ acquire an
additional phase which is integrated exponential of ${\cal A}_{{\rm F},\pm}
(x,t)$. Whereas the dynamical phase ${\gamma}_{\rm F}^{\rm dyn}$
provides information about the duration of the evolution, the
Berry's phase reflects the nontrivial holonomy of the Fock states
on ${\cal A}^1$.

However, a direct computation of the diagonal matrix elements of
$\frac{\delta}{\delta A_1(x,t)}$ in ~\ref{eq: summa} requires a
globally single-valued basis for the eigenstates $|{\rm F},A(t);\pm \rangle$
which is not available. The connections ~\ref{eq: dvatri} can be
defined only locally on ${\cal A}^1$, in regions where
$[\frac{e_{+}b{\rm L}}{2 \pi}]$ is fixed. The values of $A_1$ in regions
of different $[\frac{e_{+}b{\rm L}}{2 \pi}]$ are connected by
topologically nontrivial gauge transformations.
If $[\frac{e_{+}b{\rm L}}{2 \pi}]$ changes, then
there is a nontrivial spectral flow , i.e. some of energy levels
of the first quantized fermionic Hamiltonians cross zero and change
sign. This means that the definition of the Fock vacuum of the second
quantized fermionic Hamiltonian changes (see Eq.~\ref{eq: deset}
and ~\ref{eq: odinodin}). Since the creation and annihilation operators
$a^{\dagger}, a$ (and $b^{\dagger}, b$ ) are
continuous functionals of $A_1(x)$, the definition of all excited
Fock states $|{\rm F},A(t) \rangle$ is also discontinuous. The
connections ${\cal A}_{{\rm F},\pm}$ are not therefore well-defined 
globally.
Their global characterization necessiates the usual introduction of
transition functions.

Furthermore, ${\cal A}_{{\rm F},\pm}$ are not invariant under 
$A$--dependent
redefinitions of the phases of the Fock states: $|{\rm F},A(t);\pm \rangle
\rightarrow \exp\{-i {\chi}_{\pm}[A]\} |{\rm F},A(t);\pm \rangle$, and 
transform like a $U(1)$ vector potential
\[
{\cal A}_{{\rm F},\pm} \rightarrow {\cal A}_{{\rm F},\pm} +
\frac{\delta {\chi}_{\pm}[A]}{\delta A_1}.
\]

For these reasons, to calculate ${\gamma}_{\rm F}^{\rm Berry}$ it
is more convenient to compute first the $U(1)$ curvature tensors
\begin{equation}
{\cal F}_{\rm F}^{\pm}(x,y,t) \equiv \frac{\delta}{\delta A_1(x,t)}
{\cal A}_{{\rm F},\pm}(y,t) - \frac{\delta}{\delta A_1(y,t)}
{\cal A}_{{\rm F},\pm}(x,t)
\label{eq: dvacet}
\end{equation}
and then deduce ${\cal A}_{{\rm F},\pm}$.

i) $n$-particle states $(n \geq 3)$.

For $n$-particle states $|m_1;m_2;...;m_n; A;\pm \rangle$
$(m_1<m_2<...<m_n)$, the ${\rm U}(1)$ curvature tensors are
\[
{\cal F}_{m_1,m_2,...,m_n}^{\pm}(x,y,t) 
= i \sum_{k=0}^{\infty}
\frac{1}{k!} \sum_{\overline{m}_1, \overline{m}_2, ...,
\overline{m}_k \in \cal Z} \{ \langle m_1;m_2:...;m_n; A;\pm|
\frac{\delta}{\delta {A}_1(x,t)}
\]
\[
|\overline{m}_1;\overline{m}_2;
...;\overline{m}_k; A;\pm \rangle 
\cdot \langle \overline{m}_1; \overline{m}_2; ...; \overline{m}_k;
A;\pm| \frac{\delta}{\delta A_1(y,t)}|
m_1;m_2;...;m_n; A;\pm \rangle - (x \leftrightarrow y) \}
\]
where the completeness condition ~\ref{eq: plete} is inserted.

Since
\[
\langle m_1;m_2;...;m_n; A;\pm |\frac{\delta
:\hat{\rm H}_{\pm}:}{\delta  A_1(x,t)}|
\overline{m}_1;\overline{m}_2;...;\overline{m}_k; A;\pm \rangle
= {\hbar} \{ \sum_{i=1}^k {\varepsilon}_{\overline{m}_i,\pm} \cdot
{\rm sign}({\varepsilon}_{\overline{m}_i,\pm}) 
\]
\[
-\sum_{i=1}^n {\varepsilon}_{m_i,\pm} \cdot
{\rm sign}({\varepsilon}_{m_i,\pm}) \} \cdot
\langle m_1;m_2;...;m_n;A;\pm|\frac{\delta}{\delta A_1(x,t)}|
\overline{m}_1; \overline{m}_2;...;\overline{m}_k; A;\pm \rangle
\]
and $:\hat{\rm H}_{\pm}:$ are quadratic in the positive and negative
chirality creation and annihilation operators, the matrix elements
$\langle m_1;m_2;...;m_n; A;\pm|\frac{\delta}{\delta  A_1(x,t)}|
\overline{m}_1;\overline{m}_2;...;\overline{m}_k; A;\pm \rangle$
and so the corresponding curvature tensors
${\cal F}_{m_1,m_2,...,m_n}^{\pm}$ and Berry phases
${\gamma}_{m_1,m_2,...,m_n;\pm}^{\rm Berry}$ vanish for all values
of $m_i (i=\overline{1,n})$ for $n \geq 3$.

ii) $2$-particle states.

For $2$-particle states $|m_1;m_2; A;\pm \rangle$ $(m_1<m_2)$,
only the vacuum state survives in the completeness condition inserted
so that the curvature tensors ${\cal F}_{m_1m_2}^{\pm}$ take the form
\[
{\cal F}_{m_1m_2}^{\pm}(x,y,t) = \frac{i}{{\hbar}^2} \frac{1}
{({\varepsilon}_{m_1,\pm} \cdot {\rm sign}({\varepsilon}_{m_1,\pm}) +
{\varepsilon}_{m_2,\pm} \cdot {\rm sign}({\varepsilon}_{m_2,\pm}))^2}
\]
\[
\cdot \{ \langle m_1;m_2;A;\pm| \frac{\delta :\hat{\rm H}_{\pm}:}
{\delta A_1(y,t)}|{\rm vac}; A;\pm \rangle 
\langle {\rm vac};A;\pm|\frac{\delta :\hat{\rm H}_{\pm}:}
{\delta A_1(x,t)}|m_1;m_2;A;\pm \rangle -
(x \leftrightarrow y) \}.
\]
With $:\hat{\rm H}_{\pm}:(t)$ given by ~\ref{eq: hamil},
${\cal F}_{m_1m_2}^{\pm}$ are evaluated as
$$
{\cal F}_{m_1m_2}^{\pm}= \left \{
\begin{array}{cc} 
0 & \mbox{for $m_1,m_2 >[\frac{e_{\pm}b{\rm L}}
{2\pi}] \hspace{3 mm} {\rm and} \hspace{3 mm} m_1,m_2 \leq 
[\frac{e_{\pm}b{\rm 
L}}{2\pi}]$},\\ \mp \frac{e_{\pm}^2}{2{\pi}^2} \frac{1}{(m_2-m_1)^2}
\sin\{\frac{2\pi}{\rm L}(m_2-m_1)(x-y)\} & \mbox{for
$m_1 \leq [\frac{e_{\pm}b{\rm L}}{2\pi}], m_2>[\frac{e_{\pm}b{\rm L}}
{2\pi}]$},
\end{array}\right.
$$
i.e. the curvatures are nonvanishing only for states with one
particle and one hole.

The corresponding connections are easily deduced as
\[
{\cal A}_{m_1m_2}^{\pm}(x,t) =
-\frac{1}{2} \int_{-{\rm L}/2}^{{\rm L}/2} dy
{\cal F}_{m_1m_2}^{\pm}(x,y,t) A_1(y,t).
\]
The Berry phases become
\[
{\gamma}_{m_1m_2,\pm}^{\rm Berry} = - \frac{1}{2} \int_{0}^{\rm T} dt
\int_{-{\rm L}/2}^{{\rm L}/2} dx \int_{-{\rm L}/2}^{{\rm L}/2} dy
\dot{A}_1(x,t) {\cal F}_{m_1m_2}^{\pm}(x,y,t) A_1(y,t).
\]
If we introduce the Fourier expansion for the gauge field
\[
A_1(x,t) =b(t) + \sum_{\stackrel{p \in \cal Z}{p \neq 0}}
e^{i\frac{2\pi}{\rm L} px} {\alpha}_p(t),
\]
then in terms of the gauge field Fourier components the Berry phases
take the form
\[
{\gamma}_{m_1m_2,\pm}^{\rm Berry} = 
\mp \frac{e_{\pm}^2{\rm L}^2}{8{\pi}^2} \frac{1}{(m_2-m_1)^2}
\int_{0}^{\rm T} dt i ({\alpha}_{m_2-m_1} \dot{\alpha}_{m_1-m_2}
- {\alpha}_{m_1-m_2} \dot{\alpha}_{m_2-m_1}) 
\]
for $m_1 \leq [\frac{e_{\pm}b{\rm L}}{2\pi}],
m_2>[\frac{e_{\pm}b{\rm L}}{2\pi}]$,
vanishing for $m_1,m_2 >[\frac{e_{\pm}b{\rm L}}{2\pi}]$ and
$m_1,m_2 \leq [\frac{e_{\pm}b{\rm L}}{2\pi}]$.
Therefore, a parallel transportation of the states $|m_1;m_2;A;\pm
\rangle$ with two particles or two holes around a closed loop in
$({\alpha}_p,{\alpha}_{-p})$-space $(p>0)$ yields back the same states,
while the states with one particle and one hole are multiplied by
the phases ${\gamma}_{m_1m_2,\pm}^{\rm Berry}$.

For the Schwinger model when ${\rm N}=1$ and $e_{+}=e_{-}$
as well as for axial electrodynamics when ${\rm N}=-1$ and
$e_{+}=-e_{-}$, the nonvanishing
Berry phases for the positive and negative chirality $2$-particle states
are opposite in sign,
\[
{\gamma}_{m_1m_2,+}^{\rm Berry} = - {\gamma}_{m_1m_2,-}^{\rm Berry},
\]
so that for the states $|m_1;m_2;A \rangle =
|m_1;m_2;A;+ \rangle \otimes |m_1;m_2;A;- \rangle$
the total Berry phase is zero.
 
iii) $1$-particle states.

For $1$-particle states $|m;A;\pm \rangle$, the ${\rm U}(1)$ curvature
tensors are
\[
{\cal F}_{m}^{\pm}(x,y,t) = i 
\sum_{\stackrel{\overline{m} \in \cal Z}{\overline{m} \neq m}} 
\frac{1}{{\hbar}^2} 
\frac{1}{({\varepsilon}_{\overline{m},\pm} \cdot {\rm sign}
({\varepsilon}_{\overline{m},\pm}) - {\varepsilon}_{m,\pm} \cdot
{\rm sign}({\varepsilon}_{m,\pm}))^2}
\]
\[
\cdot \{ \langle m;A;\pm|
\frac{\delta : \hat{\rm H}_{\pm}:}{\delta A_1(y,t)}
|\overline{m};A;\pm \rangle   
\langle \overline{m};A;\pm| 
\frac{\delta :\hat{\rm H}_{\pm}:} {\delta A_1(x,t)}
|m;A;\pm \rangle - (x \longleftrightarrow y) \}. \\
\]

By a direct calculation we easily get
\begin{eqnarray*}
{\cal F}_{m>[\frac{e_{\pm}b{\rm L}}{2\pi}]}^{\pm} & = &
\sum_{\overline{m}=m-[\frac{e_{\pm}b{\rm L}}{2\pi}]}^{\infty}
{\cal F}_{0\overline{m}}^{\pm}, \\
{\cal F}_{m \leq [\frac{e_{\pm}b{\rm L}}{2\pi}]}^{\pm} & = &
\sum_{\overline{m}= [\frac{e_{\pm}b{\rm L}}{2\pi}] - m+1}^{\infty}
{\cal F}_{0\overline{m}}^{\pm},
\end{eqnarray*}
where ${\cal F}_{0\overline{m}}^{\pm}$ are curvature tensors for the
$2$-particle states $|0;\overline{m};A;\pm \rangle$ $(\overline{m}>0)$.

The Berry phases acquired by the states $|m;A;\pm \rangle$ by their
parallel transportation around a closed loop in $({\alpha}_p,
{\alpha}_{-p})$-space $(p>0)$ are
\begin{eqnarray*}
{\gamma}_{\pm}^{\rm Berry}(m>[\frac{e_{\pm}b{\rm L}}{2\pi}]) & = &
\sum_{\overline{m}=m - [\frac{e_{\pm}b{\rm L}}{2\pi}]}^{\infty}
{\gamma}_{0\overline{m};\pm}^{\rm Berry}, \\
{\gamma}_{\pm}^{\rm Berry}(m \leq [\frac{e_{\pm}b{\rm L}}{2\pi}]) & = &
\sum_{\overline{m}=[\frac{e_{\pm}b{\rm L}}{2\pi}] -m+1}^{\infty}
{\gamma}_{0\overline{m};\pm}^{\rm Berry},
\end{eqnarray*}
where ${\gamma}_{0\overline{m};\pm}^{\rm Berry}$ are phases
acquired by the states $|0;\overline{m};A;\pm \rangle$ by the same
transportation.

For the ${\rm N}=\pm 1$ models, the total $1$-particle curvature
tensor ${\cal F}_m ={\cal F}_m^{+} + {\cal F}_m^{-}$ and total Berry
phase ${\gamma}^{\rm Berry} ={\gamma}_{+}^{\rm Berry} +
{\gamma}_{-}^{\rm Berry}$ vanish.

iv) vacuum states.

For the vacuum case, only $2$-particle states contribute to the sum
of the completeness condition, so the vacuum curvature tensors are
\[
{\cal F}_{\rm vac}^{\pm}(x,y,t) = - \frac{1}{2} 
\sum_{\overline{m}_1; \overline{m}_2 \in \cal Z}
{\cal F}_{\overline{m}_1 \overline{m}_2}(x,y,t).
\]
Taking the sums, we get
\begin{equation}
{\cal F}_{\rm vac}^{\pm} =
\pm \frac{e_{+}^2}{2{\pi}} \sum_{n>0} 
( \frac{1}{2} \epsilon(x-y)
- \frac{1}{\rm L} (x-y) ).
\label{eq: dvasem}
\end{equation}
The total vacuum curvature tensor
\[
{\cal F}_{\rm vac} = {\cal F}_{\rm vac}^{+} + {\cal F}_{\rm vac}^{-}=
(1-{\rm N}^2) \frac{e_{+}^2}{2\pi} (\frac{1}{2} \epsilon(x-y) -
\frac{1}{\rm L} (x-y))
\]
vanishes for ${\rm N}=\pm 1$.

The corresponding ${\rm U}(1)$ connection is deduced as
\[
{\cal A}_{\rm vac}(x,t) = - \frac{1}{2} \int_{-{\rm L}/2}^{{\rm L}/2}
dy {\cal F}_{\rm vac}(x,y,t) A_1(y,t),
\]
so the total vacuum Berry phase is  
\[
{\gamma}_{\rm vac}^{\rm Berry} = - \frac{1}{2} \int_{0}^{T} dt
\int_{-{\rm L}/2}^{{\rm L}/2} dx \int_{-{\rm L}/2}^{{\rm L}/2} dy
\dot{A_1}(x,t) {\cal F}_{\rm vac}(x,y,t) A_1(y,t),
\]
For ${\rm N}=0$ and in the limit ${\rm L} \to \infty$,
when the second term in ~\ref{eq: dvasem} may be neglected,
the $U(1)$ curvature tensor
coincides with that obtained in  \cite{niemi86,semen87},
while the Berry phase becomes
\[
{\gamma}_{\rm vac}^{\rm Berry} = \int_{0}^{T} dt
\int_{- \infty}^{\infty} dx {\cal L}_{\rm nonlocal}(x,t),
\]
where
\[
{\cal L}_{\rm nonlocal}(x,t) \equiv - \frac{e_{+}^2}{8 {\pi}^2}
\int_{- \infty}^{\infty}
dy \dot{A_1}(x,t) \epsilon(x-y) A_1(y,t)
\]
is a non-local part of the effective Lagrange density of the CSM
\cite{sarad93}. The effective Lagrange density is a sum of the
ordinary Lagrange density of the CSM and the nonlocal part
${\cal L}_{\rm nonlocal}$. As shown in \cite{sarad93}, the effective
Lagrange density is equivalent to the ordinary one in the sense that
the corresponding preliminary Hamiltonians coincide on the constrained
submanifold ${\rm G} \approx 0$. This equivalence is valid at the
quantum level, too. If we start from the effective Lagrange density
and apply appropriately the Dirac quantization procedure, then we
come to a quantum theory which is exactly the quantum theory
obtained from the ordinary Lagrange density. We get therefore
that the Berry phase is an action and that the CSM can be defined
equivalently by both the effective action with the Berry phase
included and the ordinary one without the Berry phase.

In terms of the gauge field Fourier components, the connection
${\cal A}_{\rm vac}$ is rewritten as
\[
\langle {\rm vac};A(t)|\frac{d}{db(t)}|{\rm vac};A(t)\rangle =0,
\]
\[
\langle {\rm vac};A(t)|\frac{d}{d{\alpha}_{\pm p}(t)}|{\rm vac};A(t)\rangle
\equiv {\cal A}_{{\rm vac};\pm}(p,t)= \pm (1-{\rm N}^2)
\frac{e_{+}^2{\rm L}^2}{8{\pi}^2} \frac{1}{p} {\alpha}_{\mp p},
\]
so the nonvanishing vacuum curvature is
\[
{\cal F}_{\rm vac}(p) \equiv \frac{d}{d{\alpha}_{-p}}
{\cal A}_{{\rm vac};+} - \frac{d}{d{\alpha}_p}
{\cal A}_{{\rm vac};-} =
(1-{\rm N}^2) \frac{e_{+}^2{\rm L}^2}{4{\pi}^2} \frac{1}{p}.
\]
The total vacuum Berry phase becomes
\[
{\gamma}_{\rm vac}^{\rm Berry} = \int_{0}^{\rm T} dt
\sum_{p>0} {\cal F}_{\rm vac}(p) {\alpha}_p \dot{\alpha}_{-p}.
\]
For the ${\rm N} \neq \pm 1$ models where the local gauge symmetry
is known to be realized projectively \cite{sarad91}, 
the vacuum Berry phase is
non-zero. For ${\rm N}=\pm 1$ when the representation is unitary,
the curvature ${\cal F}_{\rm vac}(p)$ and the vacuum Berry phase
vanish.

The projective representation of the local gauge symmetry is
responsible for anomaly. In the full quantized theory of the
CSM when the gauge fields are also quantized the physical states
respond to gauge transformations from the zero topological class
with a phase \cite{sarad91}. This phase contributes to the
commutator of the Gauss law generators by a Schwinger term and
produces therefore an anomaly.

A connection of the nonvanishing vacuum Berry phase to the
projective representation can be shown in a more direct way.
Under the topologically trivial gauge transformations, 
the gauge field Fourier components
${\alpha}_p, {\alpha}_{-p}$ transform as follows
\begin{eqnarray*}
{\alpha}_p & \stackrel{\tau}{\rightarrow} & {\alpha}_p - ip{\tau}_{-}(p),\\
{\alpha}_{-p} & \stackrel{\tau}{\rightarrow} & {\alpha}_{-p} -ip{\tau}_{+}(p),\\
\end{eqnarray*}
where ${\tau}_{\pm}(p)$ are smooth gauge parameters.

The nonlocal Lagrangian
\[
{\rm L}_{\rm nonlocal}(t) \equiv  \int_{-{\rm L}/2}^{{\rm L}/2} dx
{\cal L}_{\rm nonlocal}(x,t) =
\sum_{p>0} {\cal F}_{\rm vac}(p) 
i{\alpha}_{p} \dot{\alpha}_{-p}
\]
changes as
\[
{\rm L}_{\rm nonlocal}(t) \stackrel{\tau}{\rightarrow}
{\rm L}_{\rm nonlocal}(t) - 2{\pi} \frac{d}{dt} {\alpha}_1(A;{\tau}),
\]
where
\[
{\alpha}_1(A;{\tau}) \equiv - \frac{1}{4\pi} 
\sum_{p>0} p{\cal F}_{\rm vac}(p) ({\alpha}_{-p} {\tau}_{-}
- {\alpha}_{p} {\tau}_{+})
\]
is just $1$--cocycle occuring in the projective 
representation of the gauge group. This examplifies a connection
between the nonvanishing vacuum Berry phase and the fact that the local
gauge symmetry is realized projectively.

\newpage
\section{Conclusions}
\label{sec: con}
Let us summarize.
i) We have calculated  explicitly   the Berry phase and the corresponding
${\rm U}(1)$ connection and curvature for the fermionic  vacuum and many
particle Fock states. For the ${\rm N} \neq \pm 1$ models, we get that
the Berry phase is non-zero for the vacuum, $1$-particle and $2$-particle
states with one particle and one hole. For all other many particle states
the Berry phase vanishes. This is caused by the form of the second
quantized  fermionic Hamiltonian  which is quadratic in the positive
and negative chirality creation and annihilation operators.

ii) For the ${\rm N}= \pm 1$ models without anomaly, i.e. for the SM and
axial electrodynamics, the Berry phases acquired by the negative and
positive chirality parts of the Fock states are opposite in sign
and cancel each other , so that
the total Berry phase for all Fock states is zero.

iii) A connection between the Berry phase and anomaly   becomes more
explicit for the vacuum state. We have shown  that for our model 
the vacuum Berry phase contributes to the effective action, being
that additional part of the effective action which differs it from the
ordinary one. Under the topologically trivial  gauge transformations
the corresponding addition in the effective Lagrangian  changes   by a 
total 
time derivative of the gauge group $1$-cocycle occuring in the projective
representation. This demonstrates an interrelation between the Berry
phase, anomaly and effective action.

\end{document}